\documentclass[a4paper,fleqn,usenatbib]{mnras}

\usepackage{newtxtext,newtxmath}

\usepackage[T1]{fontenc}
\usepackage{ae,aecompl}

\usepackage{graphicx}	
\usepackage{amsmath}	
\usepackage{amssymb}	

\title[Carbon \textquotesingle fluffy\textquotesingle\ interstellar dust analogs]{Carbon \textquotesingle fluffy\textquotesingle\ aggregates  produced by helium - hydrocarbon high pressure plasmas as analogs to interstellar dust}

\author[B. Hodoroaba et al.]{
Bianca Hodoroaba,$^{1}$
Ioana Cristina Gerber,$^{1}$
Delia Ciubotaru,$^{1}$
Ilarion Mihaila,$^{2}$  \newauthor
Marius Dobromir,$^{3}$
Valentin Pohoata,$^{1}$\thanks{E-mail: vpohoata@uaic.ro}
Ionut Topala$^{1}$\thanks{ E-mail: ionut.topala@uaic.ro}
\\
$^{1}$Iasi Plasma Advanced Research Center (IPARC), Faculty of Physics, Alexandru Ioan Cuza University of Iasi, \\ Blvd. Carol I No. 11, Iasi, 700506, Romania\\
$^{2}$Integrated Center of Environmental Science Studies in the North-Eastern Development Region (CERNESIM),\\ Alexandru Ioan Cuza University of Iasi, Blvd. Carol I No. 11, Iasi 700506, Romania\\
$^{3}$Research Department, Faculty of Physics, Alexandru Ioan Cuza University of Iasi, Blvd. Carol I No. 11, Iasi 700506, Romania
}

\date{Accepted XXX. Received YYY; in original form ZZZ}

\pubyear{2018}

\begin{document}
\label{firstpage}
\pagerange{\pageref{firstpage}--\pageref{lastpage}}
\maketitle

\begin{abstract}
The present study provides an investigation on the deposition of carbon-related products on different substrates using a high power dielectric barrier discharge fed with helium and saturated hydrocarbon gases. The discharge has been characterized by means of 
voltage-current measurements, while the neutral species and the dissociation compounds accumulated in the plasma reactor have been analyzed by Fourier-transform infrared (FTIR) spectroscopy. The properties of non-aromatic hydrogenated amorphous carbon (a--C:H) films and \textquotesingle fluffy\textquotesingle\ carbon dust deposited onto various substrates are analyzed by FTIR and X-ray photoelectron spectroscopy. The sp\textsuperscript{2}-hybridized fraction of a--C:H films is negligible while for the \textquotesingle fluffy\textquotesingle\ dust, does not exceed 10\%. The microscopic appearance, the hierarchical organization and the CH\textsubscript{2}~/~CH\textsubscript{3} ratio around 2 of the \textquotesingle fluffy\textquotesingle\ dust analog are results that supports the current understanding of cosmic carbon dust physico-chemical properties. 

\end{abstract}




\section{Introduction}

Infrared emission and absorption spectroscopy or imaging, applied for the study of astronomical objects, revealed many insights into the organic chemistry in space. Many satellites, space telescopes, airborne telescopes or ground based telescopes, were designed to host scientific instruments dedicated to specific wavelength ranges in near, mid and far infrared. The upcoming mission of NASA, ESA and CSA, will carry the James Webb Space Telescope (JWST) on board and after the successful deployment of the infrared telescope, the scientific community will have access to new observational data, at high resolution and may shed some new light on the hypotheses concerning the carbon compounds and their processing in space. On one hand, the presence of infrared emission bands with central wavelengths ranging from 3  $\muup$m to 12 $\muup$m (i.e. the so-called unidentified infrared bands) and related to carbon based molecules was detected for many astronomical objects. More specific, bands observed at 3.3, 6.2, 7.7, 8.6 and 11.3 $\muup$m, where initially associated with aromatic carbon vibrational features and led to the construction of polycyclic aromatic hydrocarbon (PAH) hypothesis in 1985 \citep{leger_identification_1984, allamandola_polycyclic_1985}.
On the other hand, infrared absorption spectroscopy data towards various lines of sight, Galactic or extragalactic, provided enough data to understand that, under astronomical conditions, no unique chemical composition exists for the carbon dust. Instead, carrier families based on hydrogenated amorphous carbon (HAC or a--C:H), a hypothesis launched back in 1983 \citep{duley_3.4_1983}, are considered to be responsible for the observations of specific environments, such as Galactic  center sources, dense clouds or diffuse interstellar medium. Usually, a mixture of aliphatic (e.g. 3.4, 6.9 and 7.3 $\muup$m) and aromatic absorption features are observed, the aromatic to aliphatic ratio being function on the local thermodynamic parameters, radiation fields and hydrogen content.
In order to match the astronomical observations and to elucidate the source of the unidentified infrared bands, efforts have been made either to theoretically asses the carbon materials molecular structure and spectral properties  
\citep{peeters_rich_2002, gray_modification_2004, cohen_polycyclic_2005, li_infrared_2008, draine_infrared_2008,  li_carriers_2012, papoular_carbonaceous_2013,  dhanoa_is_2014,  mauney_formation_2016}
or to design experiments for appropriate Earth based synthesis of analog materials
 \citep{ colangeli_laboratory_1997, mennella_carbon_2001, mennella_hydrogenation_2002, rotundi_production_2002, kovacevic_candidate_2005, stefanovic_hydrocarbon_2005,  dartois_iras_2007, stefanovic_plasma_2007, pino_6.2_2008, biennier_characterization_2009, carpentier_nanostructuration_2012,  gadallah_analogs_2013}.

In order to mimic the chemical and typological features of interplanetary and interstellar carbon dust, many synthesis methods have been proposed based on condensation experiments \citep{ huisken_gas-phase_2009, jager_formation_2009},  physical vapor deposition \citep{dartois_ultraviolet_2005}, combustion and pyrolysis methods 
\citep{colangeli_laboratory_1997, pino_6.2_2008, biennier_characterization_2009, carpentier_nanostructuration_2012, gadallah_analogs_2013, gavilan_vuv_2016, gavilan_polyaromatic_2017}, pulsed laser deposition \citep{mennella_c-h_2002, jager_spectral_2008, godard_ion_2011, gadallah_mid-infrared_2012, gadallah_analogs_2013}
and plasma enhanced chemical vapor deposition
\citep{colangeli_laboratory_1997, kovacevic_candidate_2005, stefanovic_hydrocarbon_2005, stefanovic_plasma_2007, mate_high_2016, molpeceres_structure_2017}.

For a long time, the plasma processes were widely used for the assisted synthesis of carbon-based products, in various allotropic forms (e.g. amorphous carbon with controllable hydrogen content, graphite, diamonds, fullerenes, graphenes, nanotubes etc.) and solid-state appearance (soot powders, thin films and thick films, coatings). Various recent technological solutions and process operational parameters can be employed in order to tune the properties of  the carbon-based material and to obtain high purity products with optimized costs: the thermal vs nonthermal plasma process, electrical excitation methods, the composition of working gas, the plasma characteristic temperatures and the substrate temperature
\citep{churilov_fullerenes_1999, dumay_structure_2002, chen_plasma_2003, hatakeyama_creation_2003, keidar_mechanism_2010, agemi_synthesis_2011, vizireanu_pecvd_2012, kang_synthesis_2013, levchenko_low-temperature_2013, sirghi_friction_2015, ichimura_multilayer_2017, vizireanu_aging_2017}.

In this respect, the plasma based synthesis became rapidly a widespread solution to obtain carbon dust and particles
\citep{shoji_spherical_2006, nagai_growth_2008, ohno_spherical_2009, pristavita_carbon_2011, al_makdessi_influence_2017, aussems_fast_2017}
or carbon interstellar dust analogs, mainly as thin films deposited onto various substrates
\citep{ kovacevic_candidate_2005, stefanovic_plasma_2007, gadallah_analogs_2013, molpeceres_structure_2017}.
Moreover, the plasma synthesized dust analogs can be used in laboratory studies concerning the effects of different physical agents on structural properties, in order to mimic astrophysical processes or observed spectral features. 
For example, the exposure of hydrogenated amorphous carbon samples to 160~nm UV photons, at doses comparable with the average doses in the diffuse interstellar medium, was carried out in order to obtain information on spectroscopic features of dust in diffuse interstellar medium such as the UV bump at 217~nm (4.6~$\muup$m\textsuperscript{-1}) \citep{gadallah_uv_2011, gavilan_vuv_2016} or the mid-far infrared bands 
\citep{mennella_uv_2001, gadallah_mid-infrared_2012, mate_stability_2014, gavilan_polyaromatic_2017}.
Moreover, exposure of hydrogenated amorphous carbon samples to an electron beam of 5~keV energy, was found to be an approach in studying the decay of the 3.4~$\muup$m band, characteristic to destruction of CH\textsubscript{3} and CH\textsubscript{2} groups in dense clouds by cosmic rays
\citep{mate_stability_2014, mate_high_2016}.
Other works were focused on ion irradiation of analogs, aiming to find explanations for the aliphatic C--H bonds destruction by cosmic rays
\citep{godard_ion_2011, dartois_swift_2017}.

Among the plasma assisted synthesis methods, the plasma generated using the dielectric barrier discharge (DBD) is not commonly used for carbon particles deposition, although the DBD technique is widely used for long time in surface treatment
\citep{borcia_dielectric_2003, liu_effects_2004, borcia_surface_2006, de_geyter_treatment_2007,  morent_non-thermal_2008, asandulesa_effect_2010},
plasma polymerization
\citep{donohoe_plasma_1979, klages_surface_2000, fanelli_deposition_2005, girard-lauriault_atmospheric_2005,  asandulesa_influence_2010,merche_atmospheric_2012,asandulesa_chemically_2013},
nanocomposite thin films deposition
\citep{jidenko_nano-particle_2007, caquineau_influence_2009, vallade_effect_2014, bazinette_atmospheric_2016, brunet_control_2017, brunet_tailored_2018},
gas conversion
 \citep{eliasson_direct_2000, liu_methane_2001, chiper_detailed_2010, obradovic_dual-use_2011, schiorlin_carbon_2016},
decontamination, sterilization or inactivation
\citep{heise_sterilization_2004, eto_low-temperature_2008, chiper_atmospheric_2011, daeschlein_skin_2012, cullen_translation_2018}
and life science applications
\citep{kuchenbecker_characterization_2009, rajasekaran_dbd_2009, brehmer_alleviation_2015, rezaei_atmospheric-pressure_2016, ito_current_2018}.
Nevertheless, some specific studies employing DBD, were performed in connection with space science
\citep{wang_interstellar_2008, thejaswini_plasma_2011, mihaila_formation_2016}.
Overall, the DBDs have attracted a significant research interest due to the flexibility to fit into more complex plasma diagnostics set-ups, their relatively standardized electrical excitation methods, their operation at atmospheric or sub-atmospheric pressure, the absence of ion etching processes, high collisionality and low Debye length, low ionization degree and low gas temperature, the pulsed chemistry and the relatively high amount of reactive species generated per unit volume.

In this work, we present a new laboratory-scale method to synthesize carbon dust analogs using the sub-atmospheric pressure plasma generated in DBD reactor fed with helium in mixture with several hydrocarbon gases (i.e. C\textsubscript{n}H\textsubscript{2n+2}, with n\,=\,1--4). The plasma assisted hydrocarbon gas conversion was successfully proved by means of gas infrared spectroscopy, either into lower mass hydrocarbons, with the same general formula or new compounds, such as C\textsubscript{2}H\textsubscript{2} or C\textsubscript{2}H\textsubscript{4}. This is a new low temperature synthesis method for carbon dust analogs in form of both, films and dust products, no other experimental observations being available using a similar plasma device.
Furthermore, we discuss the properties of hydrogenated amorphous carbon films and the \textquotesingle fluffy\textquotesingle\ carbon dust, deposited onto various substrates, from a chemical and morphological point of view.

\section{Experimental details}
\subsection{Dielectric barrier discharge reactor}

\begin{figure}
	\includegraphics[width=\columnwidth]{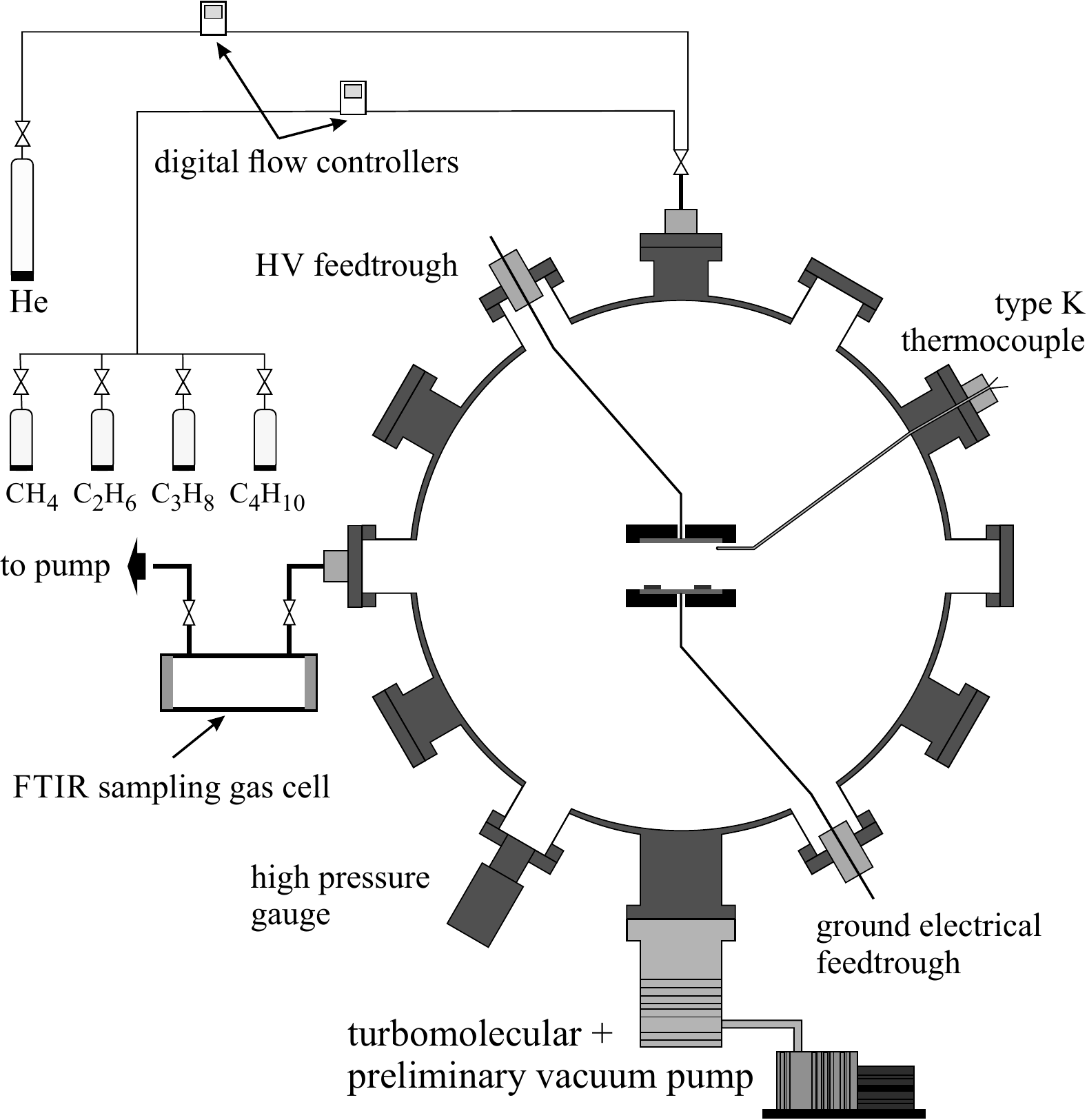}
    \caption{Schematic representation of the experimental set-up used for the synthesis of carbon-based materials.}
    \label{fig:Fig1}
\end{figure}

The carbon interstellar dust analogs were deposited on various substrates using a plasma assisted process, based on high power impulse DBD reactor, hosted by a stainless steel chamber shown schematically in Fig.~\ref{fig:Fig1}, able to generate helium-hydrocarbons plasmas at sub-atmospheric pressure (80~kPa). Prior to each deposition, proper measures were taken in order to reduce impurity contamination, i.e. atmospheric nitrogen, oxygen, carbon dioxide and water molecules adsorbed on the reactor walls. The stainless steel chamber was periodically pumped down to residual pressure of 2.6~Pa and heated up to 70~\textdegree C for several hours to allow an appropriate thermal desorption process.

The plasma is generated in a DBD assembly, fixed horizontally, composed out of two identical Aluminum electrodes (area of 43~cm\textsuperscript{2}), taped on glass dielectric plates (1~mm thickness) and separated by a 5~mm gas gap. The bottom electrode was grounded and the top electrode (i.e. the power electrode) was connected to the output of a nanosecond high voltage pulses circuit, composed of a fast transistor high voltage switch (Behlke HTS 301), backed by a DC high voltage power supply (Glassman High Voltage series KL) and triggered by a Wave Function Generator (Tektronix AFG 3022C or AF 3022B). Positive high voltage pulses of about 6~kV amplitude, 1~kHz frequency and 500~ns duration were applied on the power electrode, producing an electric field large enough to breakdown the helium-hydrocarbon gas mixture, generating an electrical discharge, confined inside the interelectrode gap. A type-K thermocouple is fixed in the middle of the interelectrode gap and its output is connected to a laboratory multimeter (UNI-T UT 71C), with the data logging optical USB communication interface enabled. The light emission from the discharge  was analyzed using an Triax 550 scanning monochromator spectrometer equipped with a CCD detector (Symphony Jobin Yvon Inc.).

Various types of substrates, such as flexible graphite (GoodFellow Co.), silicon, quartz and NaCl plates with approximately 1 cm\textsuperscript{2} in surface area, were placed directly onto the ground electrode. Gas lines are connected to the DBD reactor and the gas flow is regulated by electronic mass flow controllers (Alicat Scientific, MC Series): helium (99.996\%) and the precursor hydrocarbon gases (methane - CH\textsubscript{4}, ethane - C\textsubscript{2}H\textsubscript{6}, propane - C\textsubscript{3}H\textsubscript{8}, butane - C\textsubscript{4}H\textsubscript{10}, 99.5\%). The impurities level (ppm) for helium is H\textsubscript{2}O~<~5, O\textsubscript{2}~<~5, while for hydrocarbon gases is O\textsubscript{2}~<~100, N\textsubscript{2}~<~600, H\textsubscript{2}~<~500, other C\textsubscript{n}H\textsubscript{m}~<~3000, as stated by the producer, Linde Gas.

The DBD reactor was filled up with the gas mixture of He (85\%) - C\textsubscript{n}H\textsubscript{2n+2} (15\%), at an initial total pressure of 80 kPa, then the reactor was sealed, the gas flow shut off and the plasma was ignited for 6 hours. The applied voltage and the ground line current were sampled using a high-voltage probe (Tektronix P6015A, 1000:1, 75~MHz) and a current probe (Pearson 6485, 1~V~A\textsuperscript{-1}, 1.5~ns rise time, 200~MHz high frequency), monitored or periodically recorded using a digital oscilloscope (Tektronix TDS 5034B, 350~MHz, 5~GSa~s\textsuperscript{-1}).

Infrared spectroscopy was selected to obtain information about the neutral species existing in the DBD reactor. Gas samples were obtained before and after the 6~hours plasma operation, using a vacuumed infrared gas cell with NaCl windows, connected to the reactor chamber. Transmission spectra in the range of 3600~--~700~cm\textsuperscript{-1}, with 1~cm\textsuperscript{-1} spectral resolution, were recorded using a Bomem MB-104 Fourier transform infrared (FTIR) spectrometer.

\subsection{Characterization of carbon-based materials}

Two different types of hydrogenated carbon products are obtained in this study, as function of nucleation and growth processes on all substrates simultaneously present in the plasma during a single deposition step. On silicon, quartz and NaCl plates we obtain hydrogenated amorphous carbon thin films, while on flexible graphite  \textquotesingle fluffy\textquotesingle\ carbon dust particles are deposited. No DBD plasma sputtering was observed for all substrates used in our experiments and no deposition was observed if the hydrocarbon gases are absent from the gas mixture (i.e. if we use pure helium plasma or a gas mixture of helium (85~\%) with hydrogen (15~\%)). 
Taking this into account, different analysis techniques were chosen as function of the obtained carbon based product. The chemical characterization of the \textquotesingle fluffy\textquotesingle\ carbon dust was carried out by means of scanning electron microscopy (SEM), diffuse reflectance ultraviolet-visible (UV-Vis) spectroscopy, FTIR spectroscopy and X-ray photoelectron spectroscopy (XPS). Regarding the a--C:H thin films on NaCl, quartz and silicon substrates, the following analysis techniques were used: ellipsometry, FTIR, UV-Vis spectroscopy and XPS.

The FTIR spectra of carbon based films deposited on NaCl substrates were recorded in transmittance mode (with 4 cm\textsuperscript{-1} spectral resolution) using Bomem MB-Series 104 spectrometer. The \textquotesingle fluffy\textquotesingle\  carbon dust deposited on flexible graphite was collected and analyzed by attenuated total reflectance (ATR) technique. Due to high refractive index of carbonaceous materials, a germanium crystal was mounted on a high Signal-to-Noise-Ratio spectrometer (Jasco FT/IR-4700 with ATR-Pro One accessory).
For the FTIR transmittance film measurement, the infrared beam diameter at the sample position is 15~mm while for  ATR measurements, a 1.5~mm diameter  germanium crystal mount was used. All recorded spectra were auto-baseline corrected with a second-order polynomial function and a scale brake was introduced to exclude residual CO\textsubscript{2} contribution from  the spectrometer purge system.

The refractive indexes of a--C:H thin films were measured by principal angle-of-incidence ellipsometry technique using the reflection ellipsometer EL X-01R (632.8~nm HeNe laser).
UV-Vis spectra of a--C:H thin films deposited onto quartz substrates were recorded using a dual beam spectrometer (Thermo Scientific Evolution 300) at 1~nm resolution. The diffuse reflectance UV-Vis spectra of obtained \textquotesingle fluffy\textquotesingle\ carbon dust was measured with an integrating sphere (Avantes AvaSphere-50).

SEM observations have been made using the Quanta FEI 450 scanning electron microscope, operating at 30~kV applied on the electron gun. In order to prevent morphological change on the \textquotesingle fluffy\textquotesingle\ carbon samples, the samples were not metal coated.

The XPS spectra of carbon-based materials deposited on silicon and flexible graphite substrates were recorded with a PHI 5000 VersaProbe (PHysical Electronics) spectrometer, using a monochromatic Al K\textsubscript{alpha} X-ray source (1.486~keV), under a vacuum of about $2\times1$0\textsuperscript{-6}~Pa at a photoelectron take off angle of 45\textdegree . Regarding the spectra acquisition, the sputtering argon ion beam was not used in order to preserve the samples morphology.

\section{Results and discussions}

The main task of our study was to develop a new synthesis method, using atmospheric pressure plasmas, able to deposit carbon dust at low temperature, as analogs of the interstellar carbon grains, observed towards the galactic center or specific extragalactic objects. This section is dedicated to the comprehensive characterisation of both, the plasma phase and the deposition products.

\subsection{Helium - hydrocarbon DBD plasmas}
\begin{figure}
	\includegraphics[width=\columnwidth]{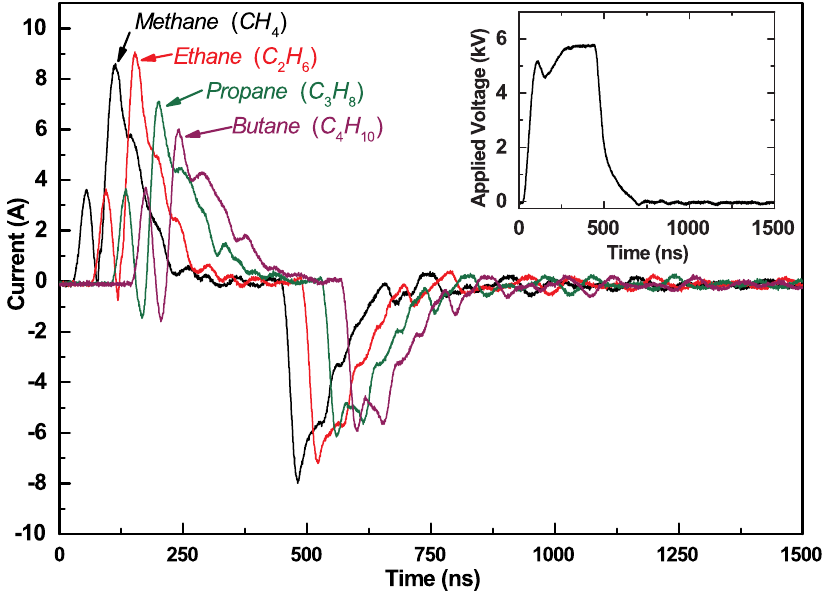}
    \caption{Typical current waveforms for the hydrocarbon containing high power impulse DBD (inset: the shape of the high voltage pulse used).}
    \label{fig:Fig2}
\end{figure}

Due to fast rise and fall time (less than 50~ns) of the high voltage pulses and the large area of DBD electrodes, the current peak (Fig.~\ref{fig:Fig2}) reaches 8~A for both, positive and negative current peaks, which corresponds to primary and secondary phases of the DBD discharges. Only slight variations  of the current peak were observed when the hydrocarbon gas type was changed. It should be pointed out here that the current, measured by the current probe, is the total current in the circuit. Mathematically, it represents the sum of the displacement current, corresponding to change oscillations inside dielectrics, the gas-gap capacitor or stray capacitors, and the conduction current, i.e. the actual discharge current. The maximum current density attains values of about 180~mA~cm\textsuperscript{-2}. Keeping constant the voltage pulse amplitude at 6~kV and replacing the methane with higher molecular mass hydrocarbons, a slight decrease of the current peak, from 8~A to 6~A, was observed. This is due to supplementary energy consumption on additional molecular fragmentation by electron collisions. A shift of the current peak position was also observed for both, positive and negative current peaks. During the 6 hours of total plasma operation time, an increase of the peak values was noticed for the measured current, of around 20\% over the initial value. 

\begin{figure*}
	\centering
	\includegraphics[width=0.8\textwidth]{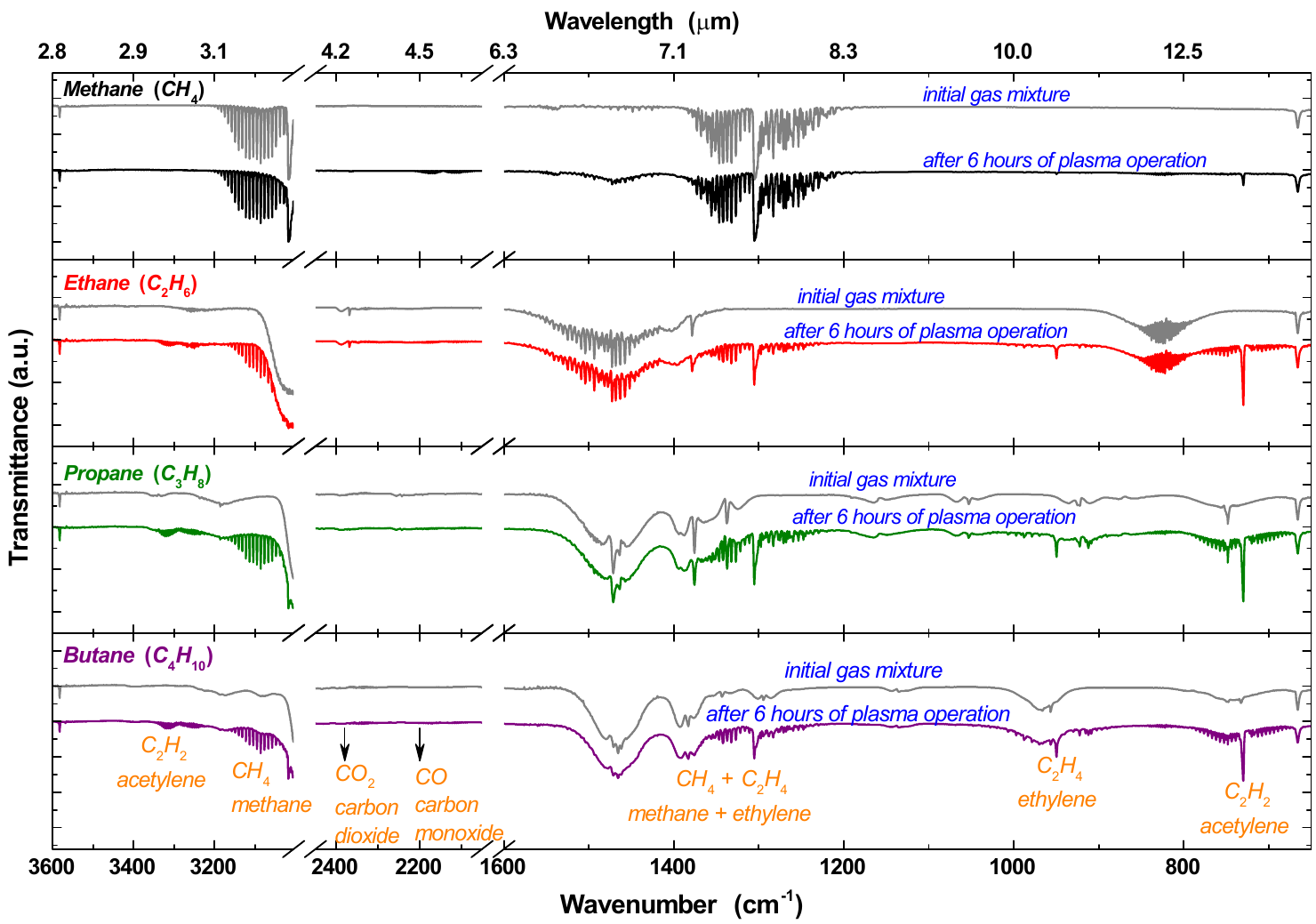}
    \caption{FTIR spectra  of the He/C\textsubscript{n}H\textsubscript{2n+2}  gas mixtures, sampled before  and after 6 hours of DBD plasma processing.}
    \label{fig:Fig3}
\end{figure*}

The analysis of the plasma emission spectra, reveals the presence of helium lines (388.8~nm, 501.6~nm, 587.6~nm, 667.8~nm, 706.5~nm, 728.1~nm), hydrogen alpha line (656.2~nm), as well carbon based radicals, the methylidyne (CH, band at 430~nm) and the diatomic carbon (C\textsubscript{2}, band at 512~nm). The gas temperature, sensed by the thermocouple increased from room temperature, at the beginning of the experiments, to a maximum value of 35~\textdegree C after approximately 100~min and then remained constant till the end of plasma operation time.

The FTIR spectra of He/C\textsubscript{n}H\textsubscript{2n+2} gas mixture obtained after plasma operation exhibits additional bands in comparison to spectra of the starting gas mixtures (Fig.~\ref{fig:Fig3}). Prominent bands assigned to acetylene, centered at 3284 and 729~cm\textsuperscript{-1},  and ethylene characteristic bands centered at 1443 and 949~cm\textsuperscript{-1} \citep{okeke_plasma-chemical_1991,deschenaux_investigations_1999,linstrom_nist_2018} were observed. The bands located at 3085 and 1305~cm\textsuperscript{-1} confirm the production of methane in the case of ethane, propane and butane containing plasmas as well. Also, we note the low level of carbon monoxide (2147~cm\textsuperscript{-1}) and the absence of carbon dioxide despite its usual strong absorption band in the infrared spectral region.

Different alkanes were utilized as precursor gases due to the fact that the main building blocks of plasma polymerization from volume fragments are sp\textsuperscript{3}-hybridized CH\textsubscript{2} and CH\textsubscript{3} functional groups. During the gas conversion, new stable precursors form in small quantities (ethylene and acetylene) meaning that some sp\textsuperscript{2} and sp-hybridized groups may contribute to the plasma deposition process.

\subsection{Hydrogenated amorphous carbon films}
The obtained plasma polymerized hydrogenated amorphous carbon (a--C:H) thin films present different thicknesses as function of the precursor gas, for an identical deposition time. The thickness increases from hundreds of nanometers to micrometers with different deposition rates varying from nearly 1 (for methane) to 3.5~nm~min\textsuperscript{-1} (for butane). Compared to other studies involving AC voltage with high frequency DBD ethylene plasma \citep{fanelli_deposition_2005}, the deposition rates for our alkane plasmas are ten times smaller. Using acetylene as gas feed precursor, \citet{mori_growth_2016}  observed that the hydrogenated amorphous carbon polymer (a--C:H) synthesized by a dielectric barrier discharge-based plasma under atmospheric pressure can be composed by micro-structured particles. The growth of the particles can be controlled by gas dilution. The a--C:H film synthesized from C\textsubscript{2}H\textsubscript{2}/N\textsubscript{2} showed cauliflower-like particles, while the film synthesized from C\textsubscript{2}H\textsubscript{2}/He was composed of domed particles. Therefore, it is reasonable to assume that the deposition of (a--C:H) films is governed by an interplay of gas phase reactions and surface processes.

\begin{figure}	
	\includegraphics[width=\columnwidth]{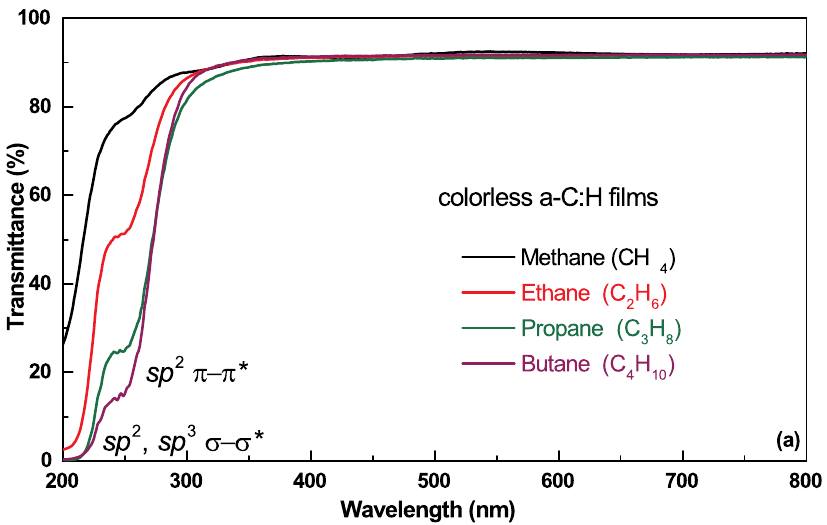}
    \includegraphics[width=\columnwidth]{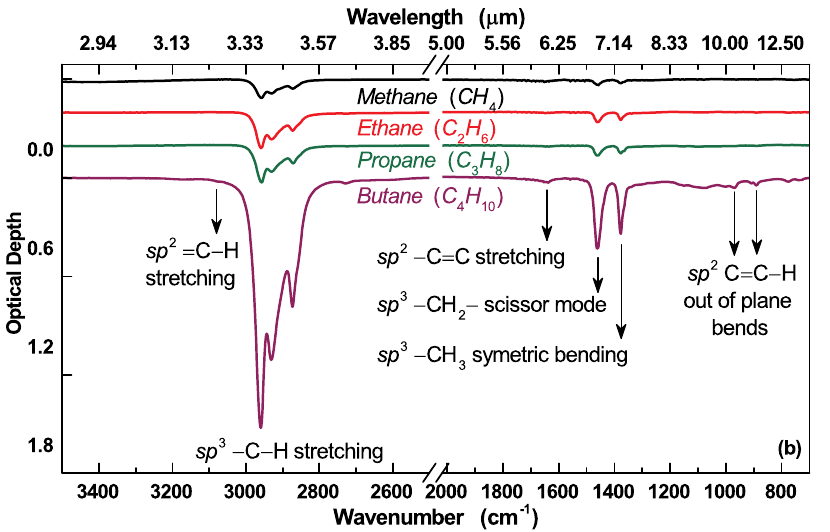}
    \caption{(a) UV-Vis and (b) FTIR spectra of hydrogenated amorphous carbon polymer thin films in hydrocarbon containing plasmas onto quartz and NaCl substrates. The optical depth scale is shifted only for better spectra visualization.}
    \label{fig:Fig4}
\end{figure}

The a--C:H thin films deposited on NaCl, quartz and silicon substrates are transparent and colorless in the visible range (Fig.~\ref{fig:Fig4}a). In the UV region absorption occurs at 250~nm (4.9~eV) related to $\piup$ bonding of carbon atoms in sp\textsuperscript{2} hybridization and at 217~nm (5.7~eV) due to sp\textsuperscript{2}, sp\textsuperscript{3} $\sigmaup$-$\sigmaup$* transition \citep{logothetidis_optical_2003, dartois_ultraviolet_2005}. 
Studies on optical and mechanical properties of amorphous hydrogenated carbon films \citep{gielen_optical_1996,logothetidis_optical_2003,dartois_ultraviolet_2005} indicate that refractive index, elasticity, hardness depends on hydrogen content and sp\textsuperscript{2}~/~sp\textsuperscript{3} ratio. More recently, from a theoretical point of view \citet{jones_variations_2012-1, jones_variations_2012-2, jones_variations_2012-3} shows that for a--C:H aliphatic materials, the CH\textsubscript{2}~/~CH\textsubscript{3} abundance ratio is correlated with hydrogen content that ranges in fixed limits.
For our a--C:H films, plasma polymerized on NaCl and quartz substrates, the refractive index has an average of 1.4 emphasizing the soft, polymer-like, character of the films, with high value of hydrogen content, and the negligible sp\textsuperscript{2} fraction.
This is consistent with previous results by Dartois et al. (2005) which observed for high aliphatic and low olefinic or aromatic networks a value of the refractive index of 1.35.

The a--C:H films obtained in DBD plasma using different feed gases were analyzed by infrared spectroscopy (Fig.~\ref{fig:Fig4}b). Specific to polymer-like amorphous carbon, common absorption bands of sp\textsuperscript{3}-hybridized carbon can be identified as: between 3000 and 2800~cm\textsuperscript{-1} (3.4~$\muup$m) due to aliphatic  $-$C$-$H stretching, 1460~cm\textsuperscript{-1} (6.85~$\muup$m) $-$CH\textsubscript{2}$-$ \ scissor mode and 1375~cm\textsuperscript{-1} (7.25~$\muup$m) $-$CH\textsubscript{3} bending mode. Small absorption peaks of carbon sp\textsuperscript{2}-hybridized are found: 3080 cm\textsuperscript{-1} (3.3~$\muup$m) due to olefinic $=$C$-$H asymmetric stretching and 1640 cm\textsuperscript{-1} (6.1~$\muup$m) due to olefinic $-$C$=$C stretching. Absorption bands specific to oxygen bonds as  1700~cm\textsuperscript{-1} (5.9~$\muup$m)  C$=$O (carbonyl group) stretching or the broadband from 3300~cm\textsuperscript{-1} (3~$\muup$m) O$-$H (hydroxyl group) stretching are insignificant, indicating a low level of oxygen contained in our a--C:H films. Similar to other results \citep{fanelli_deposition_2005,buijnsters_hydrogen_2009} in our DBD synthesized a--C:H films, hydrogen is preferentially bonded to sp\textsuperscript{3}-hybridized C atoms.

The aliphatic sp\textsuperscript{3} asymmetric modes ratio CH\textsubscript{2}~group~(2925~cm\textsuperscript{-1})~/~CH\textsubscript{3} group (2955 cm\textsuperscript{-1}) give structural information about polymerized a--C:H films. Each bonding type has a different intrinsic strength (in units of cm group\textsuperscript{-1}) and these proportionality constants were used in previous similar analyses to calculate the column density of CH\textsubscript{2} and CH\textsubscript{3}, as well the CH\textsubscript{2}~/~CH\textsubscript{3} ratio for various astronomical objects and analogs \citep{sandford_interstellar_1991, pendleton_near-infrared_1994, jacob_experimental_1996, ristein_comparative_1998, dartois_organic_2004, dartois_ultraviolet_2005, chiar_structure_2013}.

For 3.4 $\muup$m infrared band, the absorption strengths were evaluated using proportionality constant for CH\textsubscript{2} and CH\textsubscript{3}  modes and the equation 1 from  \citet{chiar_structure_2013} arising to an absorption strength ratio:
\begin{equation*}
\frac{\sigma_{CH_{2}}}{\sigma_{CH_{3}}}\approx0.62.  
\end{equation*}
Using the  equation 2 from the same study, the CH\textsubscript{2}~/~CH\textsubscript{3} ratio arise from the ratio of integrated area of each individual hydrocarbon component as follow: 
\begin{equation*}
\frac{CH_{2}}{CH_{3}}~=~\frac {Area_{CH_{2}}}{Area_{CH_{3}}}\times\frac{\sigma_{CH_{3}}}{\sigma_{CH_{2}}}~=~1.6\times\frac {Area_{CH_{2}}}{Area_{CH_{3}}}
\end{equation*}

The area ratio of asymmetric modes, CH\textsubscript{2} (2925~cm\textsuperscript{-1})  and CH\textsubscript{3} (2955~cm\textsuperscript{-1}), was determined by Gaussian fit of each 3.4~$\muup$m infrared band spectrum and mediated on different samples. The difficulty in spectral profile analysis comes from presence of a Fermi resonance peak \citep{dartois_ultraviolet_2005} and does not allow a better area ratio estimation and further we estimate a 10~\% uncertainity for CH\textsubscript{2}~/~CH\textsubscript{3} measurement from infrared spectroscopy.  For our DBD polymer-like a–-C:H films the CH\textsubscript{2}~/~CH\textsubscript{3} ratio is approximately 1, meaning that the number of methyl ($-$CH\textsubscript{3}) and methylene ($-$CH\textsubscript{2}$-$) groups are nearly equal.

The ratio of the olefinic  sp\textsuperscript{2} CH\textsubscript{2} asymmetric stretching (3080~cm\textsuperscript{-1}) to aliphatic sp\textsuperscript{3} CH\textsubscript{3} asymmetric stretching (2955~cm\textsuperscript{-1}) \citep{ristein_comparative_1998, dartois_ultraviolet_2005} indicates that our films contain no more than 5\% olefinic sp\textsuperscript{2} CH\textsubscript{2}  bonding with respect to the aliphatic sp\textsuperscript{3} CH\textsubscript{3}. The a--C:H film structure is similar to polypropylene, but presents highly branched units, random, entangled structures and incorporates also a small fraction of olefinic (alkene) hydrocarbon bridges. Our DBD polymer a--C:H film does not contain long aliphatic (alkane) chains H\textsubscript{3}C$-$(CH\textsubscript{2})\textsubscript{n}$-$ with n greater than 4 \citep{buijnsters_hydrogen_2009,gadallah_hydrocarbon_2015} and no aromatic  C$-$H or C$=$C infrared signatures were found.

\subsection{\textquotesingle Fluffy\textquotesingle\ carbon dust}
\begin{figure*}	
	\includegraphics[width=0.8\textwidth]{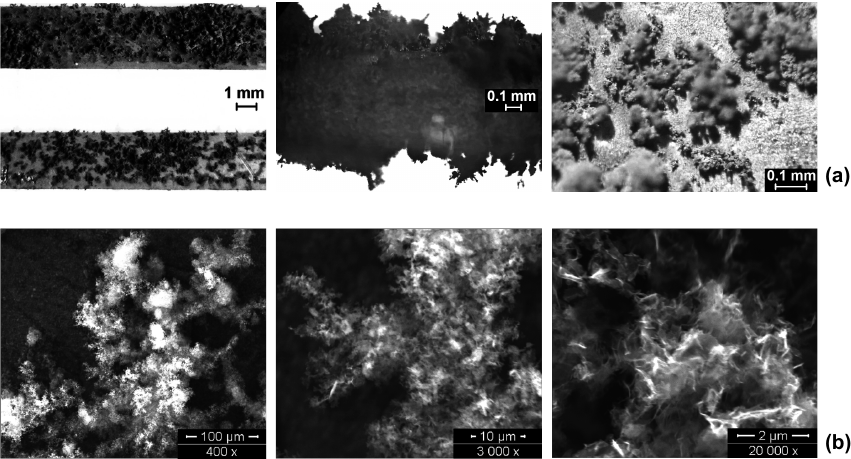}
    \caption{Island growth of \textquotesingle fluffy\textquotesingle\ carbon dust deposited on flexible graphite, in helium - propane DBD plasma (a); Typical SEM micrographs of \textquotesingle fluffy\textquotesingle\ carbon dust grains (b).}
    \label{fig:Fig5}
\end{figure*}

In contrast with smooth and totally transparent films plasma polymerized on NaCl, quartz and silicon substrates, a \textquotesingle fluffy\textquotesingle\  ash-like powder is formed on flexible graphite substrate. The deposition process is strongly influenced by the substrate nature, i.e. nucleation process on a graphite substrate appears to favor the synthesis of the \textquotesingle fluffy\textquotesingle\  aggregates of dust carbon-based material. Optical microscopy and SEM studies of all samples, performed in different substrate regions and at various magnifications, covering the mesoscale range, show various morphological features of the \textquotesingle fluffy\textquotesingle\  deposits (Fig.~\ref{fig:Fig5}). An incomplete substrate coverage was observed for all studied conditions, the coverage degree being random for different samples, but higher than 50\% in most cases (Fig.~\ref{fig:Fig5}a). The carbon-based material synthesized onto flexible graphite is composed of inhomogeneous clusters, that are due to aggregation of individual flakes with characteristic dimensions down to nano-scale (Fig.~\ref{fig:Fig5}b). A cauliflower-like growth is likely to take place during \textquotesingle fluffy\textquotesingle\  dust carbon-based material deposition, explaining the microscopic voids visible at mesoscale \citep{zani_ultra-low_2013} and the large internal surface area. 

The largely inhomogeneous structures observed in SEM images, at different magnifications, for \textquotesingle fluffy\textquotesingle\ carbon dust deposited onto flexible graphite substrates, indicates that the contribution of plasma volume reactions is insignificant and the formation of active sites that initiate and sustain the plasma polymer growth occurs at the substrate surface. This is typical to deposition process in DBD reactors, due to relatively high operating pressure and the subsequent very short mean free path.

The diffuse reflectance spectra of obtained \textquotesingle fluffy\textquotesingle\ carbon dust (reflectance less than 5\%) shows high absorption coefficient in the ultraviolet-visible-near infrared spectral range, with no significant photoluminescence effect. This result is consistent with those of \citet{godard_ion_2011}, that observed an absorption continuum due to electronic transitions for soot samples.

The refractive index for \textquotesingle fluffy\textquotesingle\ carbon dust was estimated, from the theoretical studies of  \citet{jones_variations_2012-2, jones_variations_2012-3} on physical properties of carbonaceous materials, to be around 2.

\begin{figure}	
	\includegraphics[width=\columnwidth]{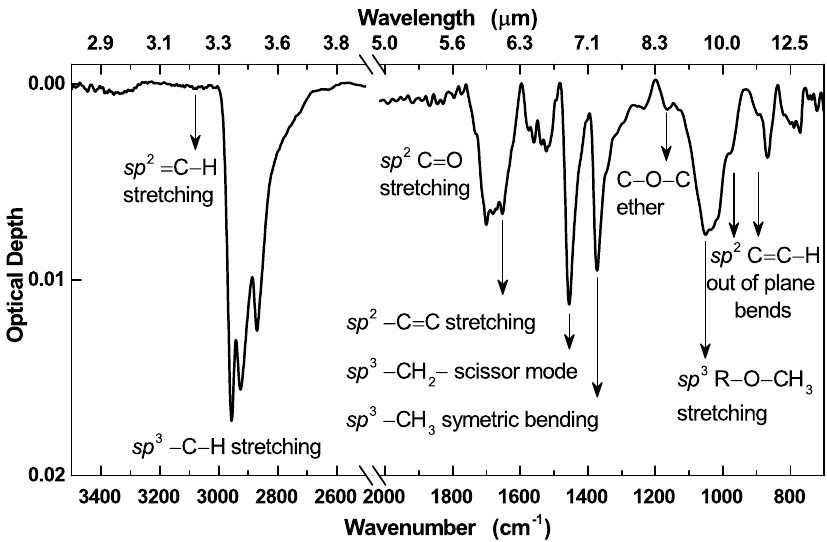}
    \caption{FTIR-ATR spectra of \textquotesingle fluffy\textquotesingle\ carbon dust produced in propane containing DBD plasma onto flexible graphite.}
    \label{fig:Fig6}
\end{figure}

A typical FTIR-ATR spectrum of the \textquotesingle fluffy\textquotesingle\ carbon dust deposited on flexible graphite in DBD plasma is shown in Fig.~\ref{fig:Fig6}. The product was collected from the flexible graphite, placed on the germanium ATR crystal, covering the entire area. Using the ATR clamp screw, the dust was carefully pressed, ensuring a sample thicker than the penetration depth of the evanescent wave at infrared wavelengths. Due to the \textquotesingle fluffy\textquotesingle\ appearance of the carbon dust, after the pressing process, air inclusions trapped in contact with the ATR crystal may still be present and diminish the absorption, resulting in a slight decrease of the optical depth values.
The spectra reveals common and additional absorption bands to a--C:H films of carbon sp\textsuperscript{3}-hybridized as: between 3000 and 2800~cm\textsuperscript{-1} (3.4~$\muup$m) due to aliphatic $-$C$-$H stretching, 1460~cm\textsuperscript{-1} (6.85~$\muup$m) $-$CH\textsubscript{2}$-$ scissor mode and 1375~cm\textsuperscript{-1} (7.25~$\muup$m) $-$CH\textsubscript{3} bending mode. Based  on the absence of the O$-$H hydroxyl group IR signature, the aliphatic ether C$-$O$-$C functional group has been identified at 1150~cm\textsuperscript{-1} (8.57~$\muup$m).  
Another ether functional group was identified at 1050~cm\textsuperscript{-1} (9.52~$\muup$m) most probably the  methoxy  R$-$O$-$CH\textsubscript{3} due to plasma hydrocarbon working gases.
Absorption bands of carbon sp\textsuperscript{2}-hybridized are found: 1700~cm\textsuperscript{-1} (5.9~$\muup$m) carbonyl C$=$O stretching and 1640~cm\textsuperscript{-1} (6.1~$\muup$m) specific to olefinic $-$C$=$C stretching. The olefinic $=$C$-$H asymmetric stretching at 3080~cm\textsuperscript{-1} (3.3~$\muup$m) is significantly small. 
Different types of alkene were identified in the fingerprint region as out of plane wagging modes of trans-alkene $-$CH$=$CH$-$ at 965~cm\textsuperscript{-1} (10.36~$\muup$m) and vinylidene $>$CH$=$CH\textsubscript{2} at 890~cm\textsuperscript{-1} (11.23~$\muup$m) \citep{dartois_ultraviolet_2005, godard_photoluminescence_2010}.
No aromatic bands at 3050~cm\textsuperscript{-1} (3.28~$\muup$m) due to $=$C$-$H stretching or 1590~cm\textsuperscript{-1} (6.28 ~$\muup$m) due to $-$C$=$C stretching were found.
The  broad peak from 1500~cm\textsuperscript{-1} (6.66~$\muup$m) corresponds to residual water vapours from FTIR spectrometer chamber and have nothing in common with the the fluffy dust.

The CH\textsubscript{2}~/~CH\textsubscript{3} ratio of \textquotesingle fluffy\textquotesingle\ carbon dust is approximately 2, using the same calculation method as in the case of our DBD a--C:H film. This is similar to the average CH\textsubscript{2}~/~CH\textsubscript{3} ratio estimated for astronomical objects in diffuse interstellar medium \citep{sandford_interstellar_1991,pendleton_near-infrared_1994,dartois_organic_2004,dartois_ultraviolet_2005,dartois_iras_2007}.

\begin{figure}	
	\centering
    \includegraphics[width=0.9\columnwidth]{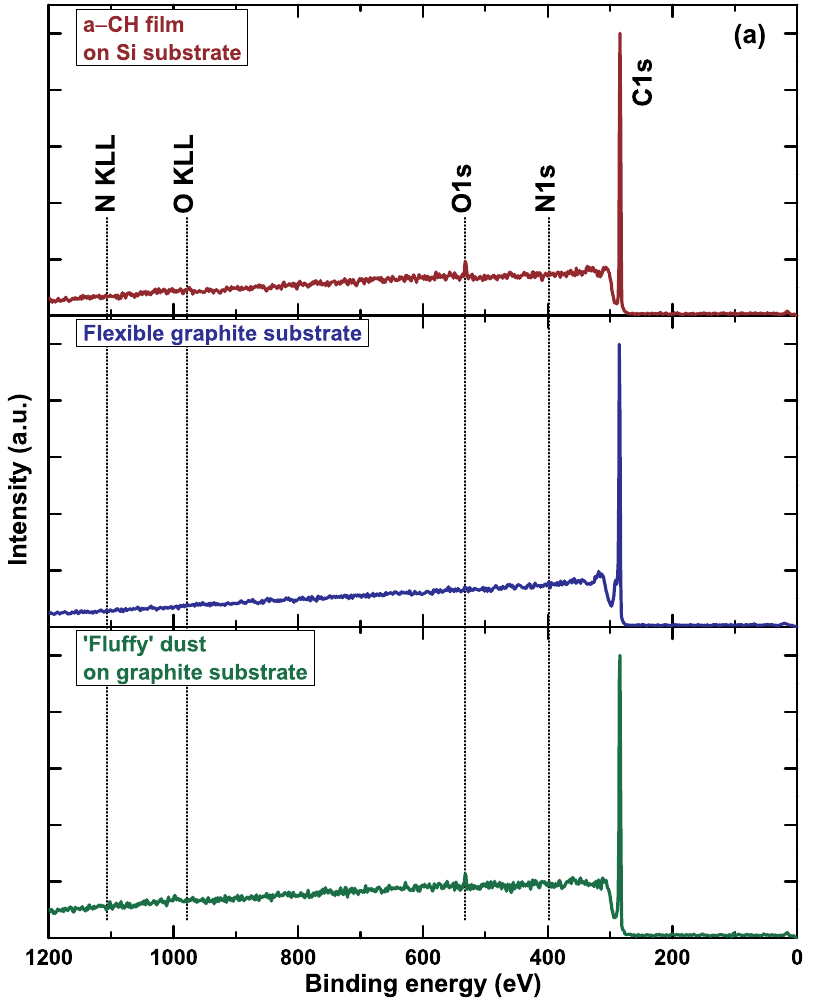}
    \includegraphics[width=0.9\columnwidth]{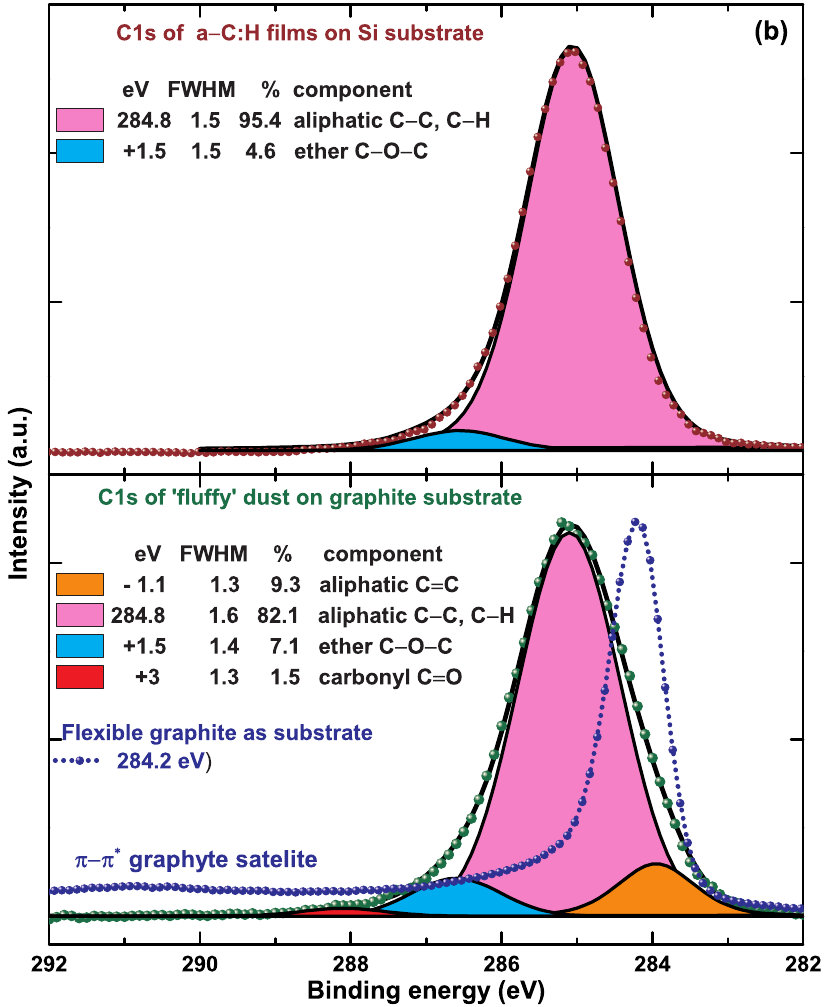}
    \caption{XPS survey spectra (a) and high-resolution scan spectra over: a--C:H film deposited on silicon substrate and  \textquotesingle fluffy\textquotesingle\ dust deposited on flexible graphite in a propane containing plasma (b).}
    \label{fig:Fig7}
\end{figure}

Switching to different hydrocarbon plasma type, chemical composition analyzed by XPS of both hydrogenated amorphous carbon (a--C:H) polymer films and \textquotesingle fluffy\textquotesingle\ carbon dust show similar results. The survey spectra shows dominant C1s peak while O1s and N1s are within detection limits (Fig.~\ref{fig:Fig7}a). 
In all experimental conditions, the carbon-based films and also the \textquotesingle fluffy\textquotesingle\ dust are not contaminated by N and O, as the C, N, O content is dominated by C at the 97\% level, whereas O can contribute to at most 2\% and N~<~1\%. 
No significant variations of relative atomic percentage as function of hydrocarbon gas type was observed during this study. The presence of oxygen indicates that the surface of dust samples is slightly oxidized (either during synthesis, either due to ambient air exposure after deposition).

High resolution XPS spectra measured on a--C:H film deposited on silicon substrate and  \textquotesingle fluffy\textquotesingle\ carbon dust deposited on flexible graphite as substrate, both synthesized in propane containing plasma were selected as example of carbon hybridization analysis. The C1s  envelopes were deconvoluted using Grams Galactic software with 90\% Gaussian - 10\% Lorentzian peak fitting mixed profile (Fig.~\ref{fig:Fig7}b). The Doniach-Sunjic asymmetric fitting function \citep{jackson_determining_1995,cunge_dry_2015} was excluded from our XPS fitting procedure due to non aromatic or non graphene-like character of our  a--C:H film and \textquotesingle fluffy\textquotesingle\ dust. As an example of asymmetric dependency, the XPS spectrum of flexible graphite was overlapped on Fig.~\ref{fig:Fig7} for comparison. The accepted binding energy for C1s peak in pure graphite is 284.2~eV ($\pm$~0.2~eV) \citep{weckhuysen_characterization_1998,filik_xps_2003,gaddam_physical_2013} with a full width at half maximum (FWHM) less than 1~eV \citep{filik_xps_2003} and a satellite peak shifted with $\sim$~6~eV to higher binding energy, specific to $\piup$-$\piup$* ring transition \citep{jackson_determining_1995,asandulesa_chemically_2013,gaddam_physical_2013}.

The main contribution to C1s envelopes for both, a--C:H film and \textquotesingle fluffy\textquotesingle\ dust (Fig.~\ref{fig:Fig7}b) was assigned to saturated aliphatic hydrocarbon sp\textsuperscript{3} hybridized, carbon atoms from C$-$C bonds and CH\textsubscript{2}  or CH\textsubscript{3}  functional groups at the binding energy of  of 285.1~eV ($\pm$~0.2~eV) \citep{weckhuysen_characterization_1998,tien_production_2011,gaddam_physical_2013}. For high concentrations of sp\textsuperscript{3} hybridized C atoms, the C1s peak tend to gain in shape symmetry and is slightly shifted towards higher binding energy in respect with sp\textsuperscript{2} carbon derived from a standard graphite. Chemical shifts of $+$1.5~eV and $+$3~eV are typically assigned to aliphatic ether C$-$O$-$C and carbonyl C$=$O functional groups \citep{weckhuysen_characterization_1998,tien_production_2011,gaddam_physical_2013,varga_diamond/carbon_2017}. Small amount of  hydrogenated aliphatic carbon,  sp\textsuperscript{2} hybridized was assigned to olefinic C$=$C bonds, as also probed by infrared analysis, at a chemical shift of $-$1.1~eV from the main sp\textsuperscript{3} hybridized carbon peak \citep{weckhuysen_characterization_1998,artemenko_amination_2015,varga_diamond/carbon_2017}.  The sp\textsuperscript{2} satellite peak specific to $\piup$-$\piup$* ring transition, is not present in main C1s peak of both, a--C:H film and \textquotesingle fluffy\textquotesingle\ dust, revealing no aromatic ring in a--C:H film and \textquotesingle fluffy\textquotesingle\ dust backbone.

XPS survey (Fig.~\ref{fig:Fig7}a) shows a low level of oxygen and nitrogen contamination. Typical infrared signatures of amine, carboxyl and hydroxyl groups are below the detection limit of the infrared spectrometer and from this point of view their contribution to  C1s envelope of a--C:H film and \textquotesingle fluffy\textquotesingle\ dust  was neglected. Nevertheless, taking into account the surface contamination processes and knowing that typical depth probed by XPS is less than 5~nm, the ether C$-$O$-$C and carbonyl C$=$O components area might include small contribution from C$-$N and O$-$C$=$O functional groups.

By combining all potential sp\textsuperscript{3} and sp\textsuperscript{2} hybridized in aliphatic carbon bonding type described by XPS analysis, the sp\textsuperscript{2} fraction for a--C:H films is negligible. In the case of the \textquotesingle fluffy\textquotesingle\ dust, the sp\textsuperscript{2} fraction does not exceed 10\%. These results are also confirmed by the infrared analysis.

\section{Astrophysical implication and conclusions}

The laboratory production of carbonaceous compounds as analogs to interstellar dust extends now over many decades and it was always of great help for astronomers concerning the composition of the interstellar dust grains. Using multiple analysis techniques, we have shown in this paper that the high power impulse dielectric barrier discharge in helium and hydrocarbon gas mixtures, is a suitable method for laboratory synthesis of \textquotesingle fluffy\textquotesingle\ aggregates, as analogs to interstellar dust grains, exhibiting both chemical and morphological similarities.

Regarding the composition of interstellar carbon dust, infrared observations of localized sources \citep{sandford_interstellar_1991,pendleton_near-infrared_1994} or along the line of sight toward Sgr~A* \citep{chiar_composition_2000}, IRAS~08572+3915 \citep{dartois_iras_2007}, IRAS~19254-7245 \citep{risaliti_revealing_2003}, NGC~1068 \citep{marco_vlt_2003} and NGC~5506 \citep{imanishi_3.4-m_2000},  confirm the presence of a solid hydrogenated carbon based material with aliphatic character identified by infrared absorption centered at 3.4, 6.85, and 7.25~$\muup$m. The FTIR spectra of both, the a--C:H films and \textquotesingle fluffy\textquotesingle\ aggregates are in a global good agreement with the above mentioned observations, i.e. for sources presenting a high column density of sp\textsuperscript{3} hybridised carbon atoms.  The bands assigned to polycyclic aromatic hydrocarbons at 3.28 and 6.28~$\muup$m \citep{chiar_near-infrared_1998, chiar_composition_2000, chiar_structure_2013}  are not present in the infrared spectra of both a--C:H films and \textquotesingle fluffy\textquotesingle\ dust.
The fraction of saturated aliphatic hydrocarbon groups and the hydrogen content are variable for all materials proposed as dust analogs. A suitable analog to interstellar hydrogenated carbon grains should have the hydrogen content that matches into the range arising from astronomical observations, as detailed by \citet{dartois_iras_2007}: from 19\% to 42\% (or hydrogen-to-carbon H~/~C ratio $\approx$ 0.23 to 0.72). Post-synthesis processing of analogs is a suitable strategy to match the above mentioned range. For example, by exposing the analogs to different external agents i.e. hydrogen bombardment \citep{mennella_hydrogenation_2002, mennella_c-h_2002}, the hydrogen content is diminishing and might be situated in the range 20\% to 42\% (H~/~C~$\approx$~0.25~$-$~0.72).

The H~/~C value is highly dependent on the relative ratio of aromatic-to-aliphatic groups in the laboratory interstellar dust analogs. For an ideal aromatic network the H~/~C goes to zero and cannot exceed 1.6 (or hydrogen content not greater than 61.5\%)  \citep{jones_variations_2012-1} for a fully aliphatic stable network. According to \citet{jacob_experimental_1996}, for a soft polymer-like film, with a density of 1.2 g cm\textsuperscript{-3}, the H~/~C ratio is around 1. The same values of density and H~/~C ratio were assumed by \citet{godard_ion_2011} for a--C:H films with the refractive index of 1.4 and later on by \citet{mate_high_2016}  for hydrogenated amorphous carbon (a--C:H) with variable mixtures of aromatic and aliphatic groups. On soot samples, polyaromatic cross-linked by aliphatic bridges, \citet{godard_ion_2011} found a higher density (1.8 g cm\textsuperscript{-3}), refractive index 1.7 and a much smaller hydrogen content (H~/~C~=~0.01) than a--C:H films. The hydrogen density is higher in the aliphatic bridges that link the aromatic units and this explains this small value of  H~/~C ratio.

Although the precise H~/~C ratio of the our a--C:H films and \textquotesingle fluffy\textquotesingle\ aggregates is not experimentally known, we may use the theoretical results of  \citet{jones_variations_2012-1} to estimate the H~/~C ratio for high aliphatic networks. For an a--C:H film with CH\textsubscript{2}~/~CH\textsubscript{3} fraction approximately 1, the hydrogen content is $\approx$~58\% (H~/~C~$\approx$~1.4). For the \textquotesingle fluffy\textquotesingle\ aggregate, the CH\textsubscript{2}~/~CH\textsubscript{3} ratio is approximately 2, that gives the hydrogen content  $\approx$~55\% (H~/~C~$\approx$~1.2). As expected for an aliphatic sample, the \textquotesingle fluffy\textquotesingle\ dust is highly hydrogenated. Considering also the 10\% of olefinic components in the \textquotesingle fluffy\textquotesingle\ dust, the hydrogen content will slight decrease \citep{godard_ion_2011} to an estimated rough value of 50\% (H~/~C~$\approx$~1).

We note now the necessity of post-synthesis processing for the \textquotesingle fluffy\textquotesingle\ carbon aggregates, in order to reach a value of the H~/~C ratio in good agreement with the astronomical observations.

Along with the infrared absorption features and the specific H~/~C ratio of dust analogs, the mesoscale appearance should be also of interest for astronomers, to match the physical properties of dust  in interplanetary space.
The analysis of dust samples obtained from nearby Earth collecting flights  \citep{messenger_opportunities_2002,  mautner_meteorite_2006,  rietmeijer_quantitative_2007}, return missions carrying appropriate dust collectors
 \citep{brownlee_comet_2006, flynn_elemental_2006,  keller_infrared_2006,  sandford_organics_2006,  westphal_interstellar_2014}
or the more recent analysis of thousands of dust grains with the on-board instruments on ROSETTA orbiter  (i.e. optical atomic force microscopes
\citep{bentley_aggregate_2016, langevin_typology_2016, ellerbroek_footprint_2017},
showed that the interplanetary, asteroidal and cometary dust particles present some common morphological features:
\begin{itemize}
\item non-regular shapes and extremely diverse size distribution;
\item variable degree of porosity and density, from compact particles (i.e non-porous, smooth, crystalline structures) to highly porous structures (i.e. loosely packed, \textquotesingle fluffy\textquotesingle\ grains);
\item a hierarchical structure of clusters or aggregates, with individual sub-micron grains, of about 500~nm or less, stacked together into larger structures, with typical dimensions from 10~microns to several hundred microns;
\item various types of clusters have been pointed (i.e. shattered clusters, glued clusters and rubble piles) based on the analysis of some morphological characteristics (e.g. the connecting matrix, if any, supposedly containing carbonaceous matter, the spatial relationships between components and their size distribution).
\end{itemize}

Some of the above mentioned morphological features of interplanetary, asteroidal and cometary dust particles, where already reproduced in laboratory dusty plasma experiments \citep{garscadden_overview_1994, szopa_pampre:_2006}.

Thus, an Earth based experiment aiming to synthesize dust analogs in general and more specifically an interstellar carbon dust analog, should address at least the following issues: the relationship between the spectral properties and the chemical structure (i.e. the aliphatic vs the aromatic content), the microscopic appearance of the product and the dust analog processing due to ionizing radiation fields or highly energetic particles bombardment.

The \textquotesingle fluffy\textquotesingle\ microscopic appearance and the hierarchical organization of the dust analog deposited onto flexible graphite substrates is consistent with the above described recent observations of interplanetary, asteroidal and cometary dust.

We conclude that the high power impulse dielectric barrier discharge in helium and hydrocarbon gas mixtures is a suitable new method for low temperature deposition of carbon dust analogs, in form of both non-aromatic thin films and \textquotesingle fluffy\textquotesingle\ aggregates. The \textquotesingle fluffy\textquotesingle\ carbon aggregates combines both, spectroscopic and morphological features of carbon cosmic dust, being suitable for laboratory mass production and ready for use in dust processing studies with radiations and particle beams.

\section*{Acknowledgements}

We would like to thank to Dr. G.B. Rusu from the USAMV, Iasi, Romania for its help during the electron microscopy imaging. Our thanks also go to Dr. R. Jijie, for her contribution during the preliminary phases of the study, and to our colleagues, Dr. G. Borcia, Dr. A. Chiper and Prof. G. Popa, for careful reading of the manuscript and helpful suggestions.

This work was supported by the Romanian Space Agency (ROSA) under the projects STAR CDI ID 486/2017~-~2018.




\bibliographystyle{mnras}
\bibliography{references} 








\bsp	
\label{lastpage}
\end{document}